\documentclass[prd,eqsecnum,apsi,twocolumn]{revtex4}

\usepackage{graphicx}
\usepackage{latexsym}
\usepackage{amsmath}
\usepackage{amssymb}

\newcommand{\be}{\begin{equation}}
\newcommand{\ee}{\end{equation}}
\newcommand{\bea}{\begin{eqnarray}}
\newcommand{\eea}{\end{eqnarray}}
\newcommand{\bfig}{\begin{figure}}
\newcommand{\efig}{\end{figure}}

\newcommand{\sech}{\hbox{sech}}
\newcommand{\cosech}{\hbox{cosech}}

\begin{document}
\title{\Large \bf Fermion localization in generalised Randall Sundrum model}
\author{Ratna Koley${}^{}$ \footnote{E-mail: tprk@iacs.res.in}, 
Joydip Mitra${}^{}$ \footnote{E-mail: tpjm@iacs.res.in} and  
Soumitra SenGupta${}^{}$ \footnote{E-mail: tpssg@iacs.res.in}}
\affiliation{ Department of Theoretical Physics and Centre for
Theoretical Sciences,\\
Indian Association for the Cultivation of Science,\\
Kolkata - 700 032, India}

\begin{abstract}

A generalized two-brane Randall-Sundrum warped braneworld model 
admits of solutions of the warp factor for both positive 
or negative cosmological constant on the visible 3-brane which can resolve the naturalness problem in connection with the fine tuning of 
 Higgs mass in the standard model of elementary particles.    
To explore the location of the standard model fermions in such a generalized warped model, we, in this work,  
determine the dependence of the localization profile of a bulk fermion on the brane cosmological constant brane tension and
the bulk fermion mass.
Our results reveal that for 
a positive and small value of the induced cosmological constant a bulk fermion is localized 
close to the brane. 
On the other hand  for a visible brane with negative cosmological constant    
and positive tension , the fermions are localized inside the bulk leading to phenomenologically interesting
possibilities.  
\end{abstract}

\pacs{04.50.+h, 04.20.Jb, 11.10.Kk}
\maketitle

\section{Introduction}
Ever since the Kaluza-Klein proposal of possible existence of extra spatial dimension(s), there have been intense activities
to explore the role of such dimensions in various observable phenomena in our four dimensional
universe. Emergence of string theory in early eighties enhanced this interest further by theoretically predicting the
inevitable presence of such extra dimensions. Furthermore some phenomenological 
extra dimensional models  were proposed in the context of the so called `naturalness problem' in the standard model of
elementary particles.    

Standard model has been  extremely successful in explaining physical phenomena up to TeV scale.
It however encounters the well-known fine tuning or naturalness problem in connection with the Higgs scalar mass. 
ADD model \cite{ADD} introduced the notion of  
large extra dimensions to bring the effective Planck scale down to TeV which removed the fine tuning problem at the expense of
introducing intermediate mass scales corresponding to the large extra dimensions {\em i.e.} which in turn brings back the hierarchy 
problem in a new guise.  
Alternatively, Randall-Sundrum warped braneworld model \cite{RS} has been particularly successful in resolving the fine tuning problem  
without bringing in any arbitrary intermediate scale between Planck and TeV. The two-brane RS model 
has the following features : 
TeV brane 
cosmological constant is zero which is close but not equal to   
the presently observed value of $\sim 10^{-48}$ $GeV^4$. In addition the TeV  brane representing our universe is possessed 
with negative brane tension.

Such braneworld scenario subsequently has been generalized such  
that the brane cosmological constant (cc) as well as the brane tension of the visible 3-brane can be 
both positive or negative \cite{bgen}. This model
also resolves the fine tuning problem but
now the values of brane-bulk parameter in the warp factor takes depends on the values of the induced cc on the 
visible brane \cite{genrs}. Such generalised scenario also admits of positive tension brane.

An important question in the context of the original RS model has been `Can the standard model fermions also propagate
in the bulk just as gravity?' From the string theory point of view the fermions being open string mode are naturally
confined to the 3-branes. However if one relaxes the string theory constraint, the fermions indeed can propagate
in the bulk. In that case , using a localizing field in the bulk ( say a scalar field) , such bulk fermion  can be localized 
at different region in the bulk by appropriately choosing  the interaction potential between the fermion and the 
scalar\cite{fermiloc, fermiloc2, fermiloc3, ref2, ref3}. Various 
phenomenological implications of having the standard model fermions inside the bulk have
been studied extensively.  It has been shown that appropriate choice for the masses of the bulk fermions determine their localization in
extra dimension which in turn leads to various interesting phenomenological consequences including
a possible  explanation for the fermion mass hierarchy in the standard model derived as an effective theory on the visible 
3-brane \cite{fermiloc, fermiloc2, fermiloc3, ref2}. Schemes of
electroweak symmetry breaking as well as chiral symmetry breaking have also been suggested by introducing a fourth generation 
fermion in the bulk \cite{ref1}.
It was also shown that a small Dirac neutrino mass can be obtained without invoking see-saw mechanism for certain
bulk fermion models \cite{ref2}. Numerous works have been done in recent years with fermions ( both massless and massive ) in the bulk 
of such warped geometric model to explore the consequent phenomenological implications \cite{bulkferm}.     

In this work, we consider a massless as well as a massive bulk fermion and study the localization profile of the fermionic 
wave function in the generalised RS model as a function of the visible brane cosmological constant for both dS and AdS cases. 
Our result demonstrates that the localization region of the bulk fermion 
crucially depends on the value as well as the sign of the induced brane cosmological constant. 
As the overlap of a fermion wave function
on the TeV brane is shown to depend  on the brane cosmological constant therefore various parameters in the 
effective four dimensional action  also in turn depends on the choice of the brane cosmological constant.
Interesting  phenomenological implications are therefore expected in such a scenario.
\section{The Model}

In the RS warped braneworld model \cite{RS} the cosmological constant (cc)  
on the visible brane is zero and also the visible brane tension is negative. 
The modulus stability of such a brane world was achieved by introducing a 
bulk scalar \cite{gw}. However, the zero cc of the visible brane 
is not consistent with the observed small value. It has been 
demonstrated in \cite{genrs} that this condition can be relaxed for a more 
general warp factor which include branes with non-zero cc \cite{bgen}. This warp 
factor has been obtained by extremising the following the action :
\begin{equation}
S = \int d^5x \sqrt{-G} ( M^3 {\cal R} - \Lambda) + \int d^4x
\sqrt{-g_i} {\cal V}_i
\end{equation}
where $\Lambda$ is the bulk cc, ${\cal R}$ is
the bulk Ricci scalar and ${\cal V}_i$ is the
tension of the $i^{th}$ brane. 
The warped geometry has been obtained by considering a constant 
curvature brane spacetime, as opposed to a flat spacetime. The general form of the warped metric is 
\be
ds^2 = e^{- 2 A(y)} g_{\mu\nu} dx^{\mu} dx^{\nu} +r^2 dy^2 \label{metric}
\ee
where  $r$ is the modulus associated with the extra dimension and $\mu , \nu$ are brane coordinate indices. The scalar mass warping is achieved through the warp factor
$e^{-A(k r \pi)}=\frac{m}{m_0}=10^{-n}$ where $k = \sqrt{ \frac{- \Lambda}{12 M^3}}$ $\sim$ Planck 
Mass, `$n$' the warp factor index must be set to $16$ to achieve the desired warping. 
Let us consider that the induced brane cc is $\Omega$. The brane metric $g_{\mu\nu}$ may correspond to dS-Schwarzschild and 
AdS-Schwarzschild spacetimes for the induced brane cc, $\Omega > 0$ and $\Omega < 0$ respectively \cite{karch}.
The magnitude of the induced cc on the brane in the generalized RS \cite{genrs} case is non-vanishing in general 
and is given by $\omega^2 = 10^{-N}$ (in Planck units) where $\omega^2 \equiv  \mp \Omega/3k^2$ for AdS and dS brane 
respectively \footnote{From now on we will use $\omega^2$ to represent induced brane cosmological constant.}.
A careful analysis reveals that for a negative brane cosmological constant N has a 
minimum value given by $N_{min}=2n$ leading to an upper bound on the magnitude of the cosmological constant while 
there is no such upper bound for the induced positive cosmological constant \cite{genrs}.

For AdS bulk ($\Lambda<0$) considering the induced cosmological constant  on the visible brane 
to be negative the following solution for the warp factor is obtained.
\be 
\label{wfads}
e^{-A} = \omega \cosh\left(\ln \frac {\omega} c_1 + ky \right)
\ee
where $\omega^2 \equiv -\Omega/3k^2$ and $c_1 = 1 + \sqrt{1 - \omega^2}$ for the warp factor normalized to unity at $y = 0$. 
One can show that real solution for the warp factor exists if and only if $\omega^2 \leq 10^{-2n}$ which leads to an upper bound for the magnitude of the cosmological 
constant ($\sim  10^{-N}$) given by $ N = N_{min}=2 n$. So, for $n=16$ one obtains $\omega^2_{max} = 10^{-32}$. Also for $N = N_{min}$ there is a degenerate solution
for ~$ kr\pi=n \ln {10}+ \ln{2}$. The two solutions of $kr \pi$ for $(N-2n)>>1$ are given by $k r_1 \pi =n \ln{10}+\frac{1}{4}10^{-(N-2n)}$ and $k r_2 \pi =(N-n)\ln{10}+\ln{4}$ which for $n=16$ and  $N=124$ becomes
\be k \pi r_1 \simeq 36.84 + 10^{-93}~,~~ k \pi r_2  = 250.07~ \ee
%
%
The hierarchy problem has been solved for the above solutions. Interestingly one can obtain the 
visible  brane tension to be positive for the second solution. Also 
the visible brane tension may be zero for $N=N_{min}=2 n$ and it is negative for  $r=r_1$. 
For the present observed value of the cosmological constant $\sim 10^{-48}$ $GeV^4$ the first solution $k \pi r_1$ corresponds to 
RS value plus a minute correction with negative tension visible brane whereas the other solution leads to a positive visible brane tension.
\bfig[h]
\includegraphics[height = 3cm, width = 5cm]{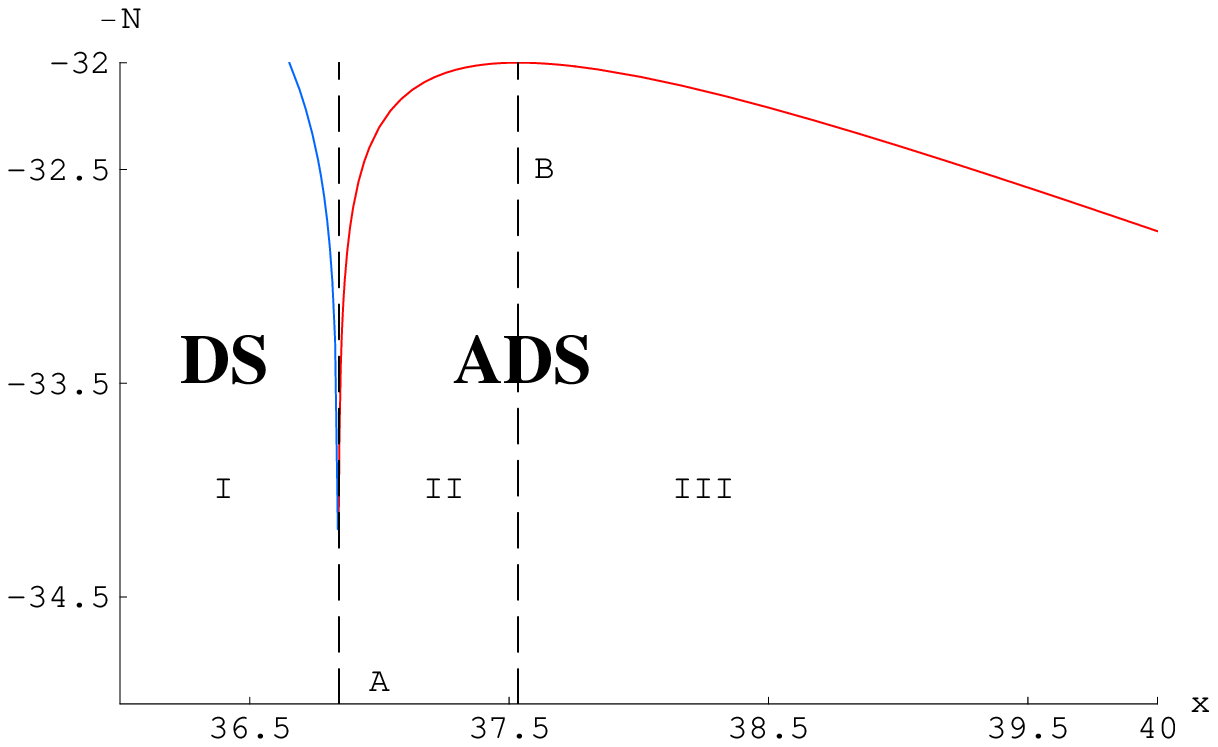}
\caption{Variation of $k r$ (which solves hierarchy problem) with brane cosmological constant.}
\label{fig:adsds}
\efig

For positive induced brane cosmological constant  the warp factor turns out to be:
\be
\label{wfds}
 e^{-A} = \omega \sinh\left(\ln \frac {c_2}{\omega} - ky \right)~
\ee
where  $\omega^2 \equiv \Omega/3k^2$ and $ c_2 = {1 + \sqrt{1
+ {\omega}^2}}~.$ In this case there is no bound on the 
value of $\omega^2$ and the (positive) brane cosmological constant can be of arbitrary magnitude. 
Also, there is a single solution of $k r \pi$ whose precise value will depend on $\omega^2$ and $n$. The result indicates that if 
one wants to resolve the Higgs mass hierarchy problem  without introducing any intermediate scale, the 
cosmological constant must be tuned to a small value close to zero.   
The variation of $k r$ (i.e. the warp factor) with 
cosmological constant (for both dS and AdS cases) as discussed above is depicted in the figure (\ref{fig:adsds}).

\section{Fermion localization}
Now we study the fermion localization scenario in the above mentioned cases.
For the general metric given in (\ref{metric})  we first find out the localization conditions and then study the different cases 
explicitly. The Lagrangian for the 5D massive Dirac fermions is given by
\be
\sqrt{-g_5} {\cal {L}}_{Dirac} = \sqrt{-g_{5}} ( \bar {\Psi} i \Gamma^a D_a \Psi - m_5 \bar {\Psi} \Psi )
\ee
where $g_5 = det(g_{ab})$ is the determinant of 5D metric and $m_5$ is the 5D fermion mass. The curved space gamma matrices are given 
by $\Gamma^a= \left ( e^{A(y)} \gamma^{\mu},-i\gamma^5 \right)$ where $\gamma^{\mu},\gamma^5$ represent 4D gamma 
matrices in chiral representation. The Clifford algebra $\{{\Gamma^a,\Gamma^b}\} = 2 g^{ab}$ is 
obeyed by curved gamma metrices \cite{fermiloc}. 
The covariant derivative can be calculated using the metric in (\ref{metric}) and is given by,
\be
 D_{\mu}  =  \partial_{\mu} - \frac{1}{2} \Gamma_{\mu} \Gamma^4 A'(y) e^{-A(y)}; ~~~~~~~
 D_5  =  \partial_y
\ee
In this  set up the Dirac Lagrangian, $\sqrt{-g_{5}} {\cal L}_{Dirac}$, turns out to be 
\be
e^{- 4 A(y)}  \bar{\Psi}
\left[i e^{A(y)} \gamma^{\mu}\partial_{\mu} + \gamma^{5} (\partial_{y} -2 A'(y)) - m_5 \right ] 
\Psi  
\ee
We decompose the five-dimensional spinor as $\Psi(x^{\mu},y)=\psi(x^{\mu}) \xi(y)$. In the 
massless case the definite chiral states $\psi_L(x^{\mu})$ and $\psi_R(x^{\mu})$  
correspond to left and right chiral states in four dimension. 
The $\psi_L$ and $\psi_R$ are constructed by, $\psi_{L,R} = \frac{1}{2} (1 \mp \gamma^5) \psi$. 
Here $\xi$ denote the extra dimensional component of the fermion wave function. 
We then can decompose five-dimensional spinor in the following way \cite{fermiloc2,fermiloc3}  
\be
 \Psi(x^{\mu},y)=\psi_L (x^{\mu}) \xi_L(y) + \psi_R (x^{\mu}) \xi_R(y)
\ee
Substituting the above decomposition in the Dirac Lagrangian we obtain the following equations for the fermions,
\be
\label{fermeql}
 e^{-A(y)}\left [\pm (\partial_{y}-2 A'(y)) + m_5 \right ] \xi_{R,L}(y)
 = - m_n ~\xi_{L,R}(y) 
\ee
The 4D fermions obey the canonical 
equation of motion, $i \gamma^{\mu} \partial_{\mu} \psi_{L,R} = m_n  \psi_{L,R}$. Because of the extremely tiny value  of the 
observed brane cosmological constant we have used near-flat 3-brane metric while evaluating the fermion equations of motion and
have kept terms only up to $\omega^2$ in the resulting expressions.
Moreover the  
equation (\ref{fermeql}) will be obtained  provided the following normalization conditions are satisfied.
\begin{eqnarray}
\label{norm1}
\int_{0}^{\pi} e^{-3 A(y)} \xi_{L_{m},R_{m}} \xi_{L_{n},R_{n}} dy = \delta_{m n} \\
\int_{0}^{\pi} e^{-3 A(y)} \xi_{L_{m}} \xi_{R_{n}} dy = 0
\end{eqnarray}
Now we study the fermion zero mode (\textit{i.e.} $m_n = 0$) localization scenario 
of both massless and massive 5D bulk fermions in the 
generalized RS model \cite{genrs} with both positive or negative brane cosmological constant. 
Consider a massless bulk fermion in the AdS 3-brane model. For the warp 
factor given in Eq. (\ref{wfads}), the left and right chiral massless modes turn out to be 
\be
\xi_{L, R} (y) = N_{1} \sech ^2 \left(\ln \frac{\omega}{c_{1}} + k y \right)
\ee

As discussed earlier, in the AdS case
there are two solutions for the modulus for a given value of brane cosmological constant which solves the hierarchy problem. 
Choosing the magnitude of $\Omega$ around $10^{-124}$ (in Planck units) we first consider the modulus values given 
by $k r_1$ around $11.73 + 10^{-93}$ which corresponds to  the negative tension visible brane. We have plotted $\xi_{L, R} (y)$ for 
this case in Fig.(\ref{fig:ads1}) for different values of cosmological constant.
\bfig[h]
\includegraphics[height = 3cm, width = 4.5cm]{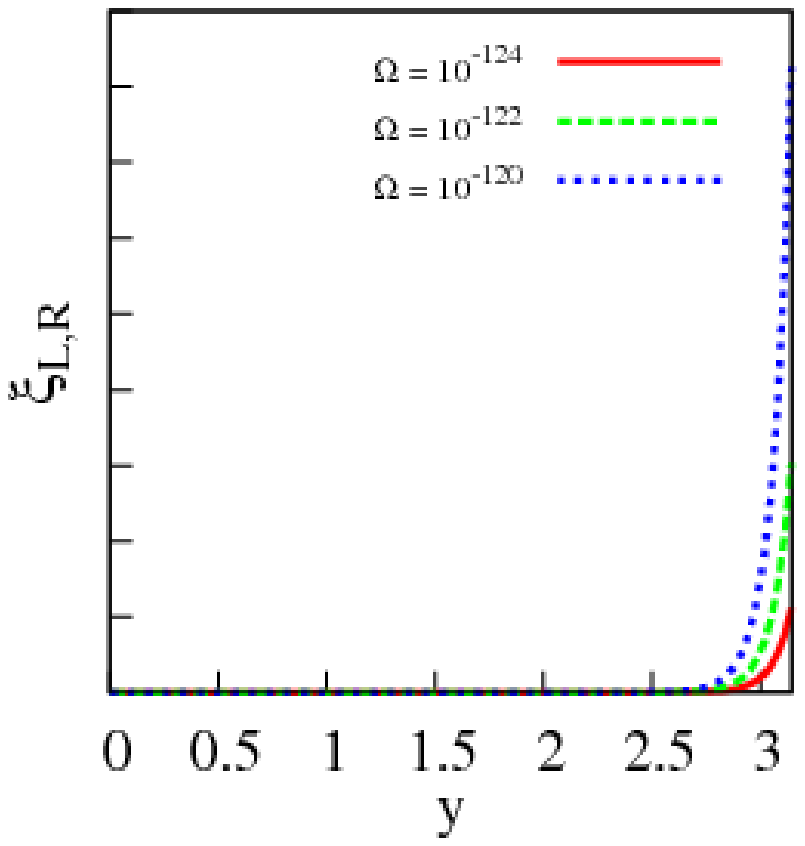}
\caption{Chiral massless modes of 5D \textit{massless} bulk fermions plotted along extra dimension for a \textit{negative tension} visible brane with \textit{negative brane  cosmological constant}.}
\label{fig:ads1}
\efig
It is apparent from the figure that as the magnitude of the induced brane cosmological constant 
becomes close to $10^{-124}$ from a higher value, masslass fermions get more and more localized on the visible brane. 
Our result reveals that smaller is  the value of the cosmological constant, 
localization of the massless fermions becomes more sharp near the negative tension visible brane which 
resembles to the confinement of open string modes to the brane.
\bfig[h]
\includegraphics[height = 3cm, width = 4.5cm]{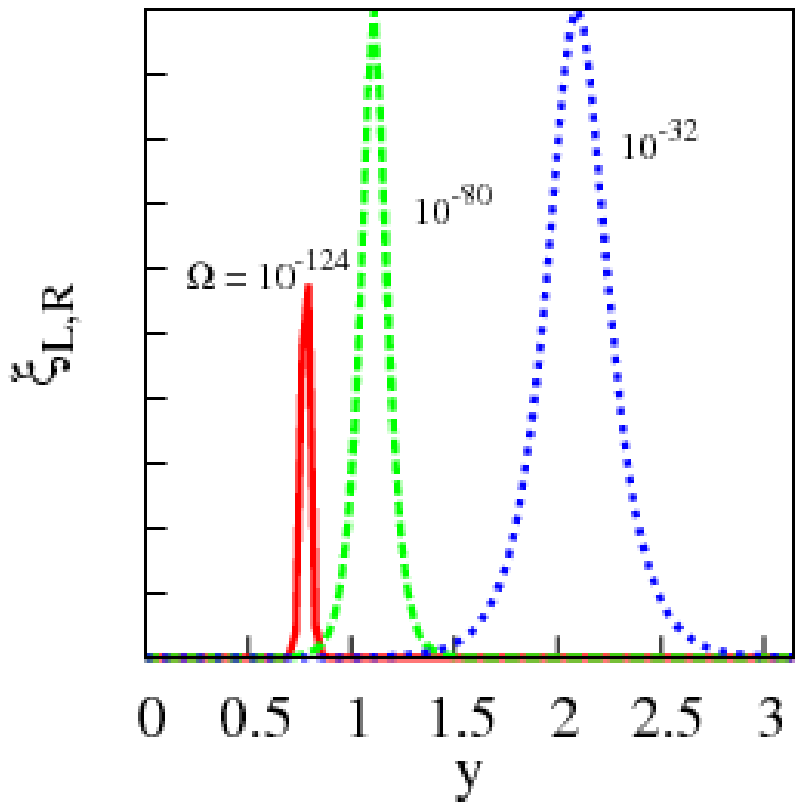}
\caption{Chiral massless modes of 5D \textit{massless} bulk fermions plotted along 
extra dimension for a \textit{positive tension} visible brane 
with \textit{negative induced cosmological constant}.}
\label{fig:ads2}
\efig

Things change drastically when we  consider the other values of the modulus 
around $kr_2\pi \sim 240$ for the same values of the cosmological constant.
In this case the visible brane tension is positive. Fig. (\ref{fig:ads2}) clearly depicts that the massless fermions in this case
are localized much away from the brane and as we shift more and more towards the negative tension visible brane i.e towards the other 
modulus value $kr_1\pi$,
by changing the value of the cosmological constant, the peak of the localization shifts towards the visible brane.
 
On the other hand for the entire range of positive values of the tension of the visible brane  
the fermions are clearly localized deep inside the  bulk space-time. 
Thus without invoking any bulk mass terms for the fermions we can localize the fermions at different region inside
the bulk by adjusting the brane cosmological constant. The corresponding
overlap of the fermion wave function with the visible brane naturally changes with the value
of the brane cosmological constant. This feature is expected to have interesting phenomenological implications
as has been discussed previously. 
Moreover a crucial point to note is that the present model  exhibits the possibility
of localizing the fermion in different region inside bulk for different choices of the brane cosmological constant
only when the brane tension is positive. Such a situation is known to be favoured 
from the point of view of stability of the brane as opposed to the branes with negative tension.
%
\bfig[htb]
\includegraphics[height = 3cm, width = 4.5cm]{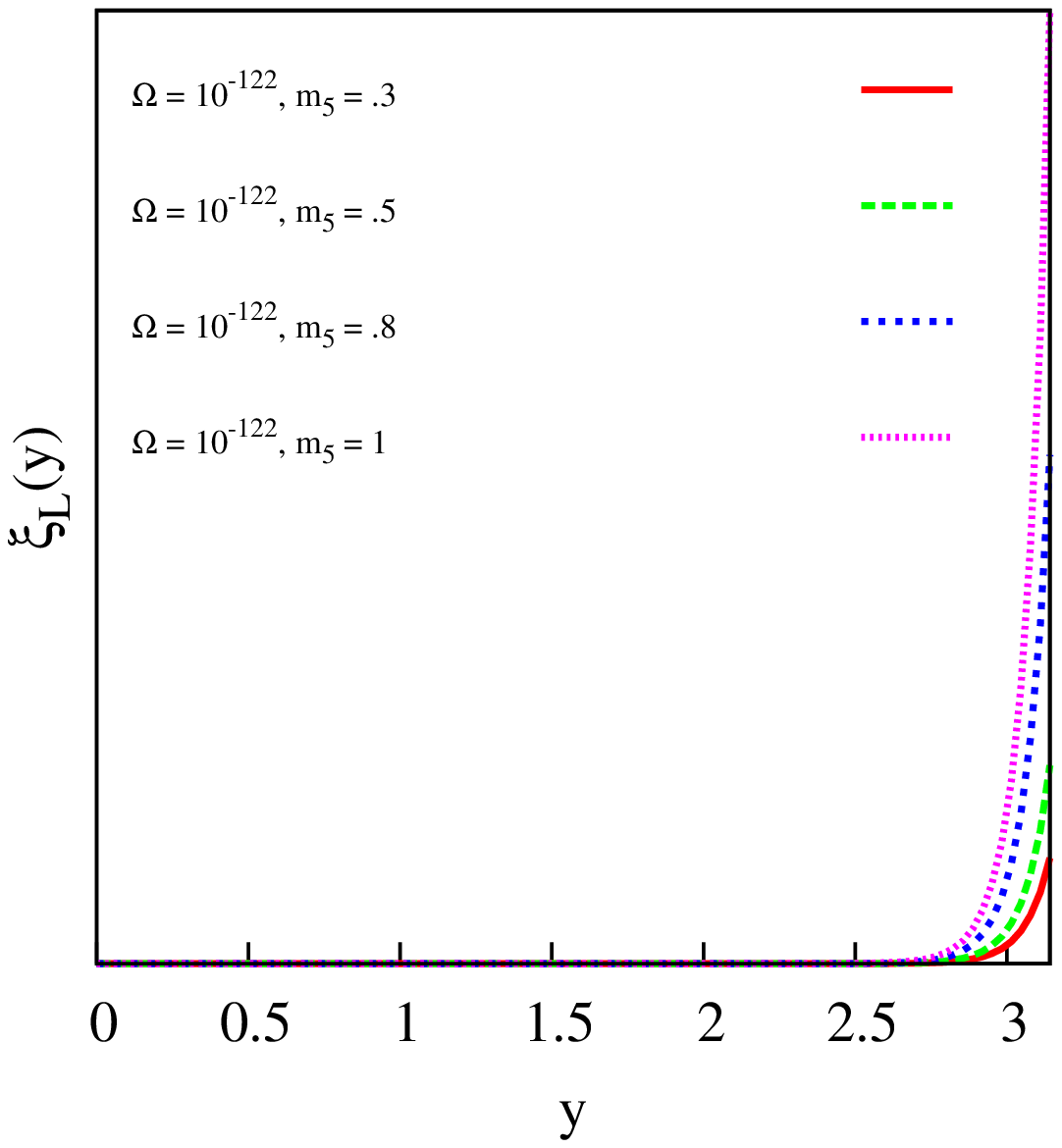}
\caption{Chiral massless modes with different 5D mass terms are plotted along extra dimension for a \textit{negative tension} visible brane with \textit{negative induced cosmological constant}.}
\label{fig:ads3}
\efig

We have also obtained interesting results considering 5D fermion 
mass to be non-zero. The profile of fermion zeromode gets further scaled by the bulk mass term, $m_5$, as follows 
\be
\xi_{L, R} (y) = N_{3} ~\sech ^2 \left(\ln \frac{\omega}{c_{1}} + k y \right) e^{\pm m_5 y}
\ee
Note that the degeneracy between the two chiral modes has been lifted by the 
mass term. For a given small value of cc as the bulk fermion becomes more and more massive 
the left chiral mode has higher peak values on the visible brane (Fig. \ref{fig:ads3}) whereas the right chiral mode 
shows exactly the reverse nature. If we consider the 5D mass term to be $c~k$ then the fermion 
wavefunction becomes $\xi_{L, R} (y) \sim {e^{(\pm c + \frac{1}{2}) k y}}/{\sqrt{1 + \frac{\omega^2 e^{2 k y}}{c^2_1}}}$. 
For $c > \frac{1}{2}$ the right handed zero mode has a very small wave function on the 
visible brane (this property helps one to obtain small neutrino masses) and these are typically 
localized near the UV brane. As the present value of cosmological constant is very small it does have much 
significant contribution to the overlap of the wave function. We further 
notice that for smaller magnitude of the cosmological constant the wave function shifts 
towards the Planck brane. It is also worthwhile to mention that the orbifold symmetry allows only one solution to be present.


We now discuss the localization scenario with massless 5D fermions on branes 
with positive brane cosmological constant.   
The corresponding warp factor in Eq. (\ref{wfds}) yields the following solutions 
for the left and right chiral massless modes of the fermions.
\be
\xi_{L, R} (y) = N_{2} ~ \cosech^2 \left( \ln \frac{c_2}{\omega} + k y \right)
\ee
\bfig[h]
\includegraphics[height = 3cm, width = 4.5cm]{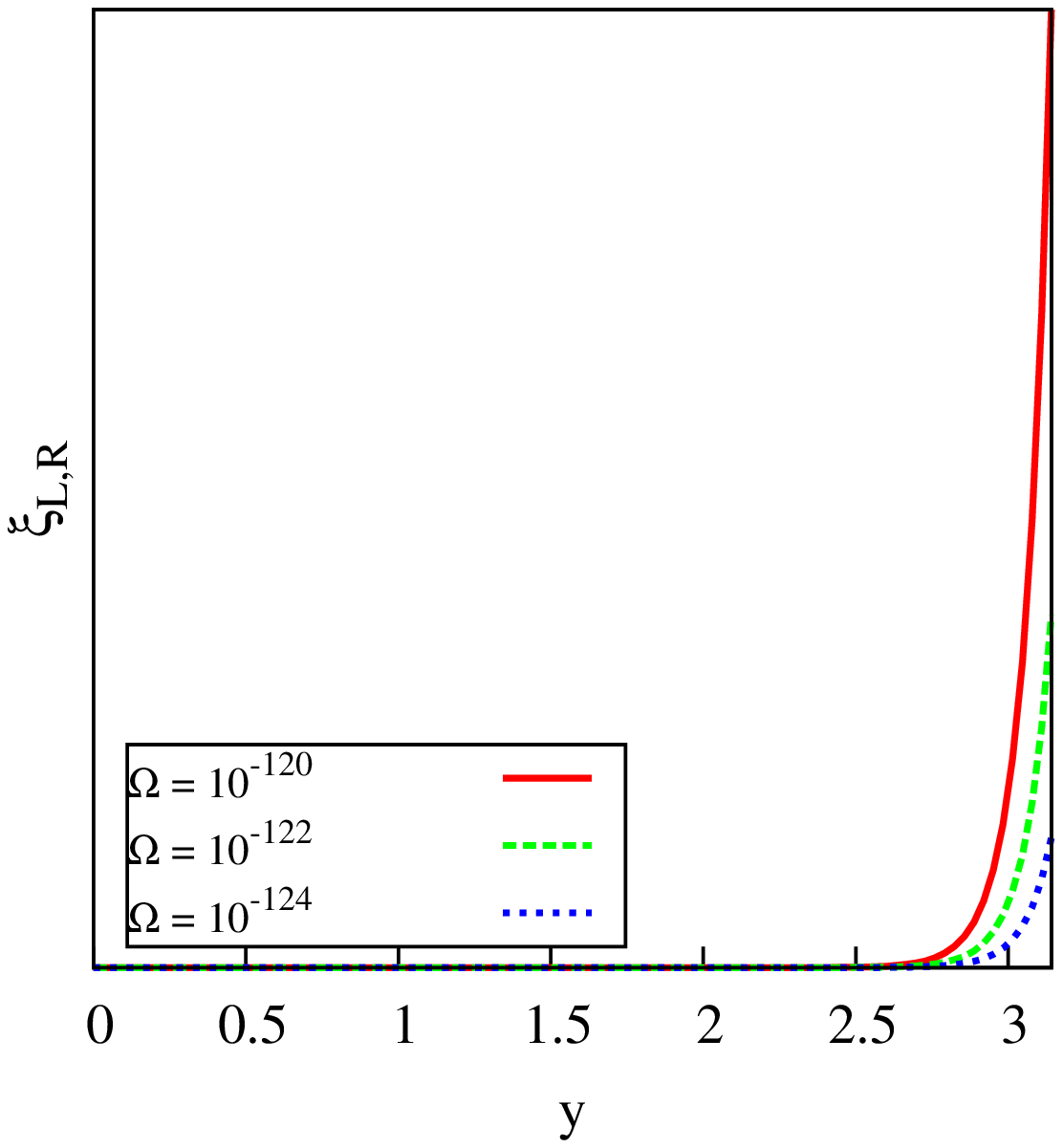}
\caption{Massless chiral fermions plotted along extra dimension for the \textit{negative tension} brane with \textit{small positive} induced cosmological constant.}
\label{fig:ds1}
\efig
The zeromodes are plotted in Fig.(\ref{fig:ds1}) for different values of the cosmological 
constant on the brane. As the value of the cosmological constant becomes smaller the 
fermions become more sharply localized on the visible brane. Thus the fact that our visible universe has a very 
small and positive cosmological constant 
indicates that the bulk fermions
are likely to get localized near the visible brane just as predicted in 
string theory. 
This in some sense, is an anthropic explanation of the 
smallness of the cosmological constant \cite{weinberg} in connection with the brane localization of the fermion in string based model.
However in this region of positive brane cosmological constant the brane tension is always negative 
and the fermion wave function peak shifts more and more inside the bulk as the value of the positive cosmological constant increase. 
This is depicted in the figure (\ref{fig:ds}). 
\bfig[htb]
\includegraphics[height = 3cm, width = 4.5cm]{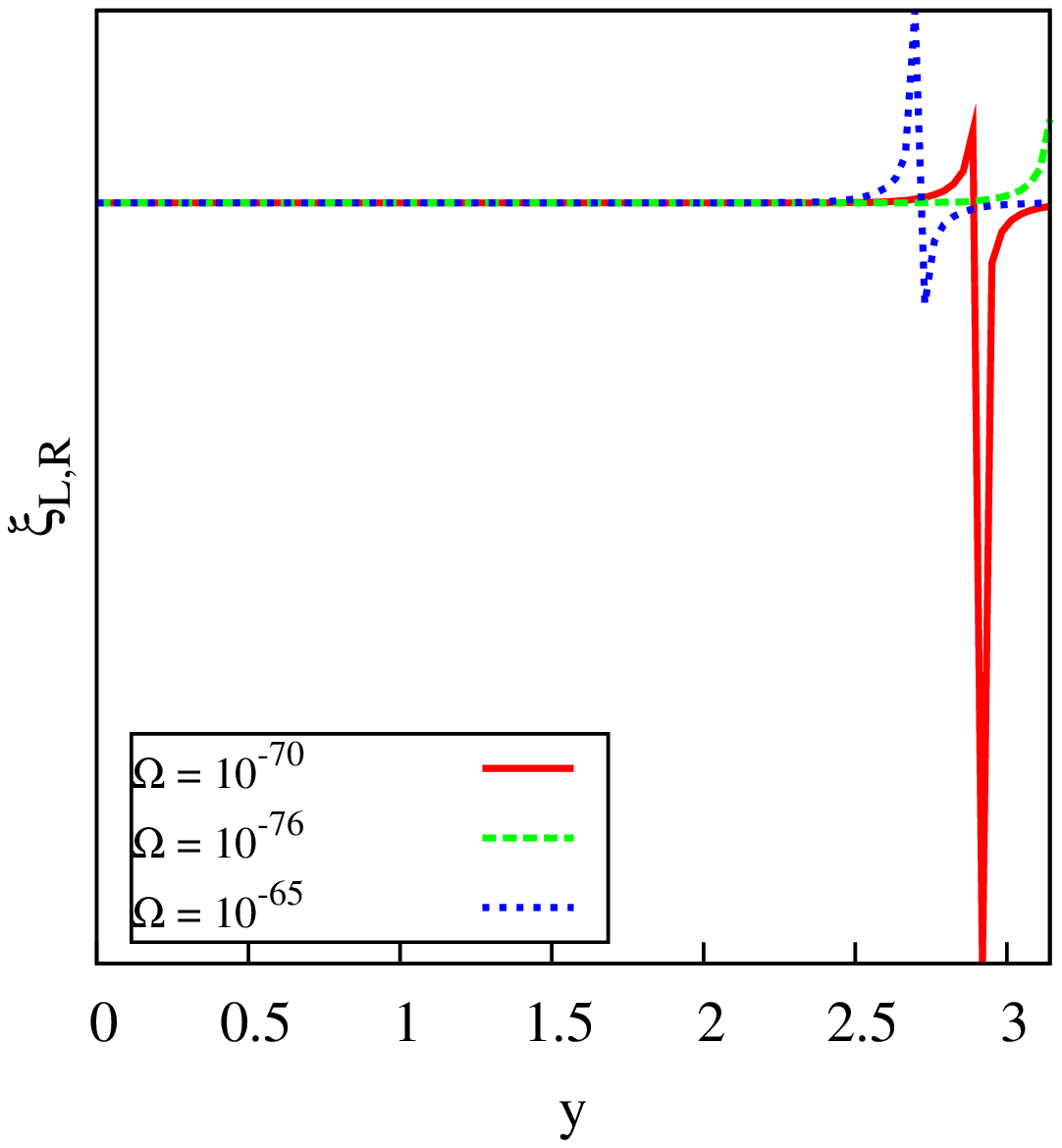}
\caption{Massless chiral fermions plotted along extra dimension 
for the \textit{negative tension brane} with \textit{large positive} induced cosmological constant.}
\label{fig:ds}
\efig
If we turn on the 5D bulk fermion mass then the fermion 
zero modes become $\xi_{L, R} (y) = N_{2} ~ \cosech^2 \left( \ln \frac{c_2}{\omega} + k y \right) e^{\pm m_5 y}$. 
For the smaller values of cc as the 5D mass term increases the left modes have higher peak value on the visible brane 
whereas the right mode gets more and more supressed.  
\bfig[htb]
\includegraphics[height = 3cm, width = 4.5cm]{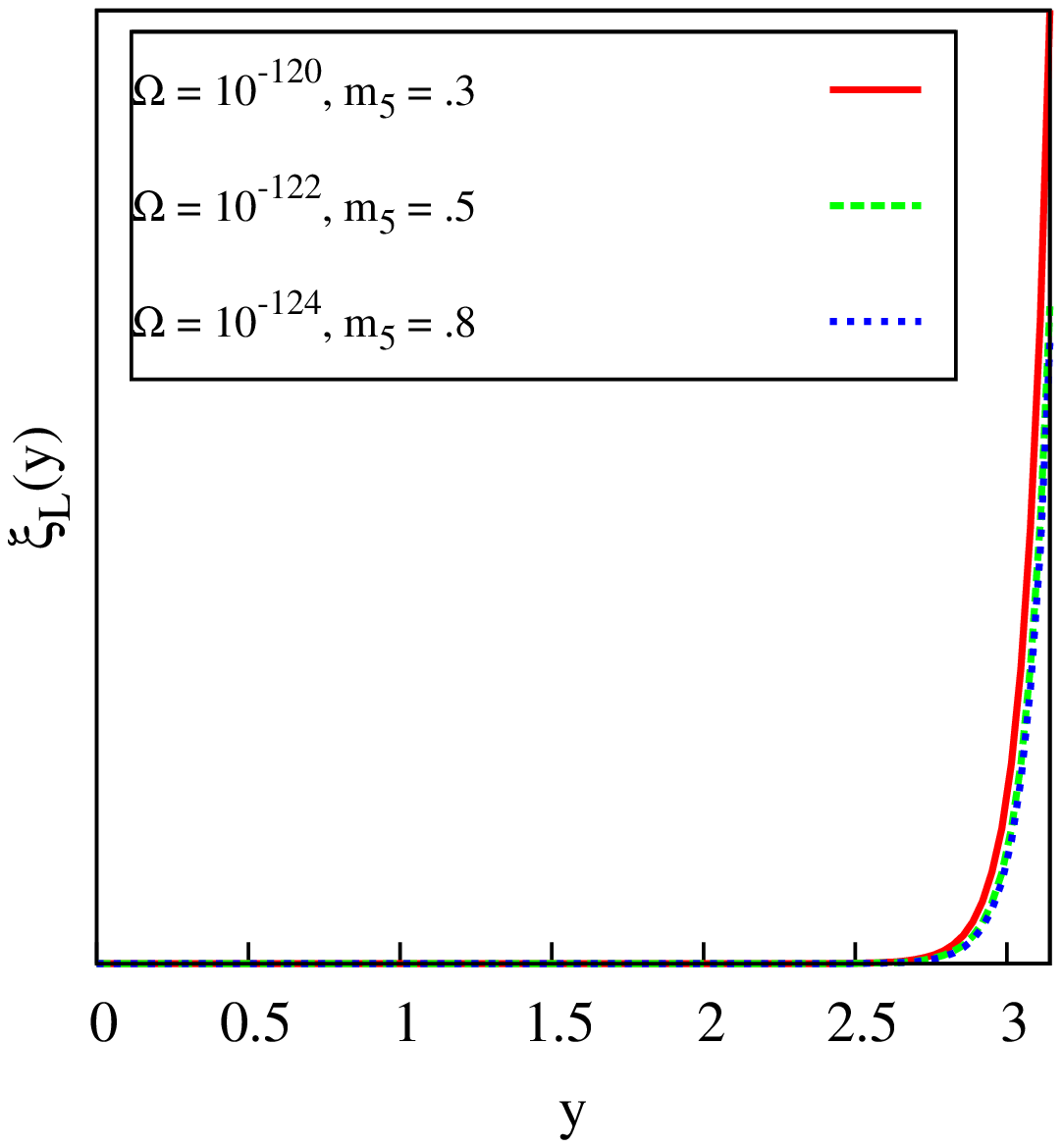}
\caption{Zeromodes of chiral fermions plotted along extra dimension for the \textit{negative tension brane} with \textit{positive} induced cosmological constant and 5D massive bulk fermion.}
\label{fig:ds}
\efig
If we consider the mass term as $c~k$ then we find that the right handed mode gets extremely supressed 
at the visible brane while getting localized near the Planck brane. This feature can give rise to the fermion mass hierarchies on the brane.


\section{Conclusion}
The localization of bulk fermions in warped brane world models have always been an 
interesting area of study specially from phenomenological
point of view. While a  string-based braneworld model prefers to have the standard model fermions being tied to
the visible brane as an open string excitation, the prospects of the presence of standard model fermions in the bulk
in much wider class of braneworld models have drawn special attention because of their important
phenomenological implications. All these models essentially deals with zero cosmological constant on the visible brane
i.e with flat 3-branes.
In this work we have determined the role of brane cosmological constant on 
the profile of a massless bulk fermion wave function and have explored how does it depend on the magnitude
as well as signature of the induced cosmological constant on the visible brane in a generalised RS model.
Our results indicate that both for de-Sitter as well as anti-de-Sitter visible brane, the localization of
the fermion wave function in the bulk becomes sharply peaked near the brane as the magnitude of the brane cosmological constant
becomes smaller. This corresponds to a string-based braneworld picture where the fermions are attached to the branes as open
string modes. On the other hand for anti-de Sitter brane, the region where brane tension is positive, we find that the fermions
are localized inside the bulk. For de-Sitter brane on the other hand, we find that  cosmological constant
can be arbitrarily large ( unlike the anti deSitter case) and the fermion wave function peaks inside the bulk
for a large cosmological constant. In both the cases the overlap
of the fermion wave function with the visible brane  depends on the value of the brane cosmological constant, bulk fermion mass and all the
other parameters in the effective 3+1 dimensional theory on the visible brane.
Thus apart from the bulk fermion mass, the brane cosmological constant also controls the effective theory on the visible 3-brane  
This implies that the brane cosmological constant has an important role in determining various phenomenological parameters
as perceived in a 3-brane slice, located at an orbifold fixed point. 

\acknowledgements{JM acknowledges CSIR, Govt. of India for providing financial support. We thank S. Roy for useful discussions 
in preparing the manuscript.}


\end{document}